# Dynamic response of HTS composite tapes to pulsed currents

V Meerovich[1], V Sokolovsky[1], L. Prigozhin[2] and D. Rozman[1]
[1]Physics Department, Ben-Gurion University of the Negev, P. O. Box 653, Beer-Sheva 84105, Israel
[2] Department of Solar Energy and Environmental Physics, Blaustein Institute for Desert Research, Ben-Gurion University of the Negev, Beer-Sheva 48990, Israel
E-mail: victorm@bgu.ac.il

**Abstract.**
Dynamic voltage-current characteristics of an HTS Ag/BiSCCO composite tape are studied both experimentally and theoretically. The tape is subjected by pulsed currents with different shapes and magnitude and voltage traces are measured using the four-point method with different location of potential taps on the sample surface. Clockwise and anticlockwise hysteresis loops are obtained for the same sample depending on location of the potential taps. The dynamic characteristics deviate substantially from the DC characteristic, especially in the range of low voltages where a criterion for the critical current value is usually chosen (1-10 μV/cm). The critical current determined from dynamic characteristics and its change with the pulse magnitude depend on location of the potential taps and on the curve branch chosen for the critical current determination (ascending or descending).
The theoretical analysis is based on a model of the magnetic flux diffusion into a composite tape for a superconductor described by the flux creep characteristic. Numerical simulation based on this model gives the results in good agreement with the experimental ones and explains the observed peculiarities of the dynamic characteristics of HTS composite tapes. The difference between the magnetic diffusion into a tape and a slab is discussed.

## 1. Introduction

Study of voltage-current characteristics (VCC) of HTS materials is of both fundamental and technological interest. It is of particular importance to reveal mechanisms responsible for dissipation and their relation to the vortex motion in superconductors. The VCC determines the transport properties of a superconductor and the possibility of its application in various devices. Pulsed current tests, allowing to prevent heating due to dissipation at the contacts, are widely used for obtaining VCCs of superconducting samples with high critical currents [1-8]. The dynamic VCCs obtained by the pulse technique provide the important information on non-equilibrium and transient processes in superconductors. Recent studies revealed various interesting phenomena associated with non-stationary vortex motion such as history (memory) effects [9, 10], different response to alternating and DC currents, effect of current sweep rate [11-13], etc. To explain these effects, a number of mathematical models were used such as a superconducting glass model, modified flux flow and resistive-weak link models, various flux creep models, etc. [14, 15].

A number of papers reported about various hysteresis loops in dynamic VCCs and related these loops to different phenomena as the inhomogeneity of the flux pinning, the phase state of the vortex matter, heating, etc. (see [9] and references 1-13 therein). The simplest explanation of the anticlockwise loops is heating. However, these loops can also result from the dynamical metastability of the vortex matter in single crystals [16]. The inelastic quasi-particle scattering can also lead to either clockwise or anticlockwise loops, depending on the magnetic field strength [17]. It was observed that the path tracing direction depended on the experimental conditions (zero-field cooling or field cooling, or field direction). In addition, the direction is influenced by inhomogeneity of the flux pinning strength for polycrystalline superconductors [18]. Thus, the character of the dynamic VCCs can be determined by different phenomena in superconductors and depends on the experimental conditions.

Despite the fact that a number of the papers was devoted to dynamic VCCs of tapes, the theoretical consideration was mostly based on the study of magnetic diffusion into an infinite slab with strongly non-linear electric field-current density (E-J) characteristics [9, 10, 19-24]. One of the conclusions resulted from this model is that an anticlockwise loop cannot be observed in a homogeneous sample in the frame of the flux creep mechanism [9]. The obtained results were usually employed to explain the experimental data obtained for samples of other shapes [13,18].

The theoretical study performed in [12, 25-27] shows that the character of the magnetic diffusion into superconducting tapes, strips and films differs sufficiently from the diffusion into an infinite slab. The correct determination of the relation between dynamic VCCs, properties of a superconductor and experimental conditions should be performed based on the models taking into account the actual shape of a sample. The problem becomes still more complicated for composite and coated superconductors, where a superconductor is shunted by a normal conducting metal.

In the present paper, we study the transient response of a monofilamentary Bi-2223/Ag tape subjected to a pulsed current. We demonstrate that the non-linear magnetic diffusion plays an important role even for relatively long current pulses and can explain many phenomena observed in the dynamic VCCs.

## 2. Experimental study
### 2.1. Experimental samples

The investigated samples of a single-filamentary Bi-2223/Ag tape were 24 mm in length, 0.2 mm in thickness and 3 mm in width. The tape was fabricated by the usual powder-in-tube method [28,29].

Two pairs of potential taps were soldered to each sample, one pair was placed on the central axis of the tape, and another was fixed close to the edge (Fig. 1). The distance between the taps belonging to the same pair was about 10 mm.

The VCC of one of the samples tested under DC conditions is shown below, in Fig. 3, (along with dynamic VCCs). This characteristic is well fitted by the power law function $E = E_0(I/I_0)^n$ with $n = 4$, $E_0 = 80$ $\mu$V/cm and $I_0 = 10$ A. This dependence is typical for the flux creep regime.

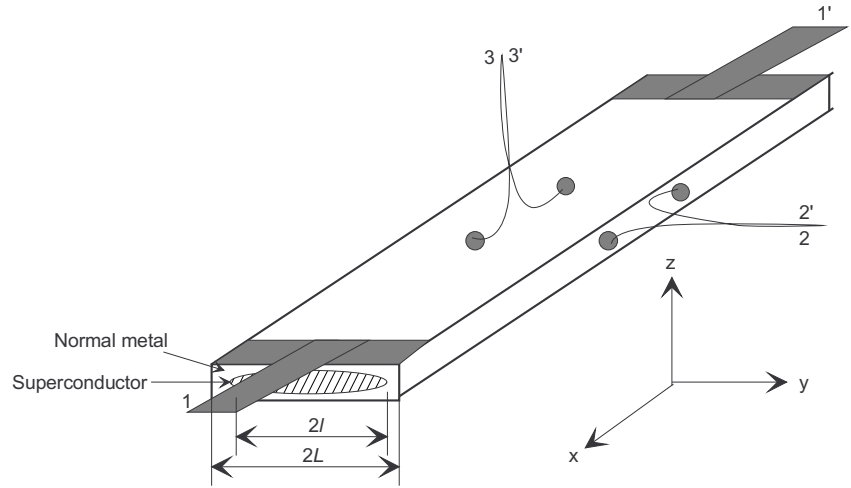

Fig. 1. A sketch of a typical experimental sample. 1-1′ - current terminals, 2-2′ and 3-3′ - wiring to edge and central potential taps.

### 2.2. Experimental procedure

We used two types of a pulse impact: (a) a relatively long pulse in the form of a half-wave of sinus typical for pulse tests; (b) a stepwise pulse with a small rise time of the current allowing us to study approaching to the stationary regime.

a) *Sinusoidal pulse*

Single current pulses of a controlled amplitude $I_0$ were obtained by a discharge of a capacitor in a circuit containing a tested superconducting sample, an inductance, and a resistor of 0.11 Ohm inserted in series. Such set-up is widely used in pulsed experiments to study VCCs of superconductors with high critical currents and/or in high magnetic fields [1, 2, 5]. The shape of the pulse was close to a half-wave of sinus, $I=I_0 \sin(\omega t)$ at $0<\omega t<\pi$, (Fig. 2a), the rise time was 0.46 ms.
The current was determined from the voltage drop across the resistor. The voltage generated across the sample was amplified by a factor of 1000 and recorded, along with the current curve, by a multichannel data acquisition device with the recording rate up to 500,000 samples/sec per channel. The measurements were performed at the temperature 77 K (in liquid nitrogen bath) for different amplitudes of the pulse.

b) *Stepwise pulse*

The tested sample was subjected to a stepwise current turned on by an electronic switch in a DC circuit containing the sample. The current front was well fitted by the function $I = I_0[1-exp(-t/\tau)]$ where $\tau = 2 \cdot 10^{-5}$ s independently of the current amplitude. The equipment described above was used for registration of the current and voltage drop from the potential taps.

*2.3. Data processing*

The voltage obtained from a pair of the taps contains the transient component induced by the changing magnetic flux in a loop formed by the potential wires and a virtual internal connection between the taps going along the tape axis [1, 2, 30]. This component contains two parts: the internal part induced by the magnetic flux linked to the inside of the sample and the outer part related to the magnetic flux that does not interact with the tape. The last voltage part induced by the magnetic flux in the measurement loop should be removed from the measured signal. Usual procedure for separation of the "useful" signal is the comparison of the voltage traces recorded at a low current, when the resistive component may be neglected, and the voltage measured above the critical current. The signal at the low current is multiplied by the ratio of "high" to "low" currents and the obtained value is subtracted from the signal at the "high" current.

This procedure is valid if the following conditions are fulfilled. First, the resistance appeared in the superconducting sample should be small enough in comparison with the resistance of the discharge circuit and the shape of the applied pulse is not distorted. Second, the configuration of the test circuit should be unchanged from measurement to measurement.

An additional problem is related to non-linear properties of a superconductor. The part of the voltage induced by the penetrating flux inside the body of the tape is strongly non-linear versus current and, therefore, the described linear subtraction procedure produces distortions of voltage curves. To decrease these distortions, one should use longer pulses or choose the "low" current as low as possible. In any case, these distortions have to be taken into account when experimental results are compared with theoretical ones. Note that the voltage registered by a pair of the taps soldered in the central points, placed on the line of symmetry, does not include the part related to the internal magnetic flux and, therefore, is not distorted by the subtraction procedure.

*2.4. Experimental results*
*2.4.1. Sinusoidal current pulse*

Fig. 2 shows the typical voltage traces for different amplitudes of the sinusoidal current pulse and demonstrates the subtraction procedure. Fig. 3 presents the dynamic VCCs of the tape obtained after application of the subtraction procedure to the measurements from the edge-assigned and central taps, respectively. The DC VCC is also shown in the same figures. All the dynamic curves demonstrate hysteresis loops. The curves obtained from the edge taps have broader loops with the intersection points and change of the path-tracing (Fig. 3a). In Section 3.3 we will show that the intersections appear as a result of the subtraction at the data processing. In the lower part (at the voltage less than about 50 µV/cm), the rising part of the curve lies significantly above the DC VCC; the return part drops down the DC curve. In the higher part, the dynamic curves lie below the DC one. The VCCs obtained with the central taps have anticlockwise loops with no intersections and lie below the DC curve (Fig. 3b).

Thus, clockwise and anticlockwise loops were observed experimentally with the same monofilamentary BSCCO tape. Simple estimates show that, in our case, these loops and the difference between the dynamic and DC characteristics cannot be explained by heating. The path-tracing of the obtained loop depends on the location of the potential taps on the sample surface. This is a characteristic property of tapes, films and other samples of planar geometry as opposed to slabs and cylindrical samples.

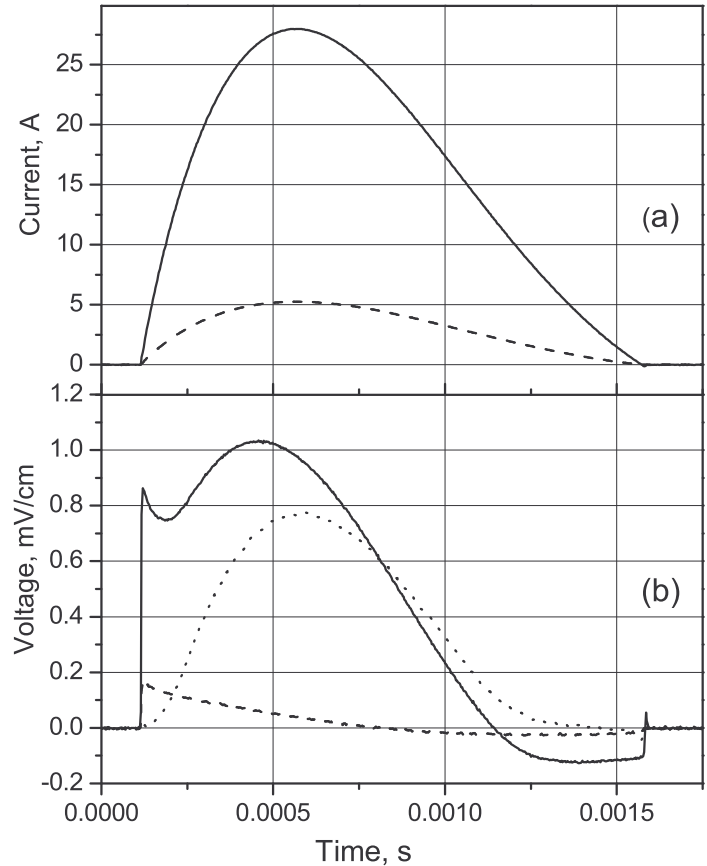

Fig. 2. Typical experimental results and illustration of the data processing procedure: (a) current pulses with the amplitude above (solid curve) and below (dashed curve) the critical current; (b) voltage traces across the potential taps at a high current (solid curve), at a low current (dashed curve) and the voltage obtained as a result of the subtraction procedure (dotted curve).

*2.4.2. Stepwise pulse*
When a stepwise pulse is applied, voltage traces contain large peaks related to the voltage induced by the fast change of the magnetic flux (Fig. 4). These peaks were removed by the subtraction procedure described above, and the resulting scope traces of the voltage on the central taps are shown in Figs. 4 and 5. One can see the delay in the appearance of the voltage signal (Fig. 5). This delay decreases with the increase of the current amplitude. This behavior shows the dependence of the characteristic time of diffusion on the current amplitude. Fig. 6 illustrates the substantial influence of the magnetic diffusion on the dynamic VCC.

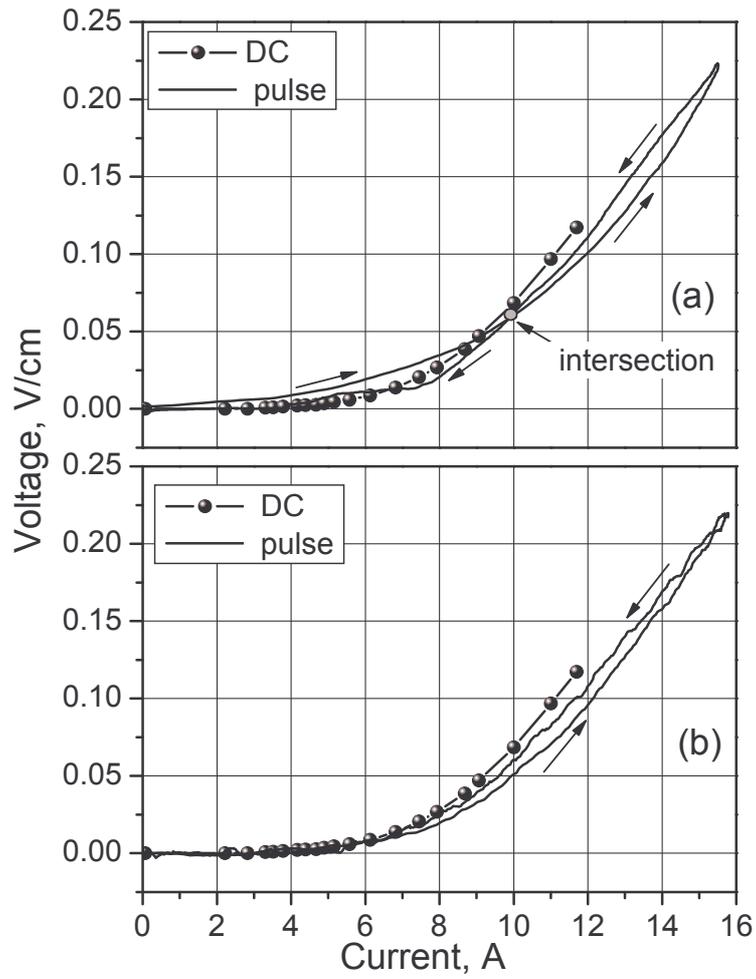

Fig. 3. Dynamic VCCs measured by the edge-assigned taps (a) and by the center-assigned taps (b). DC VCC is shown by the curve with solid circles. The arrows indicate the path tracing. The shape of the current pulse is shown in Fig. 2a.

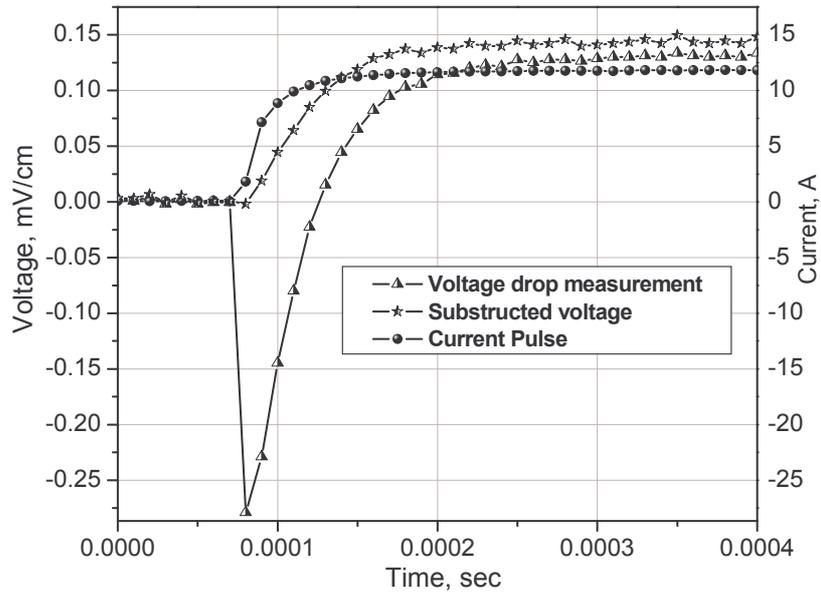

Fig. 4. Response to a stepwise current: shown the current pulse, the measured voltage drop and the voltage after processing (the subtraction of an induced component).

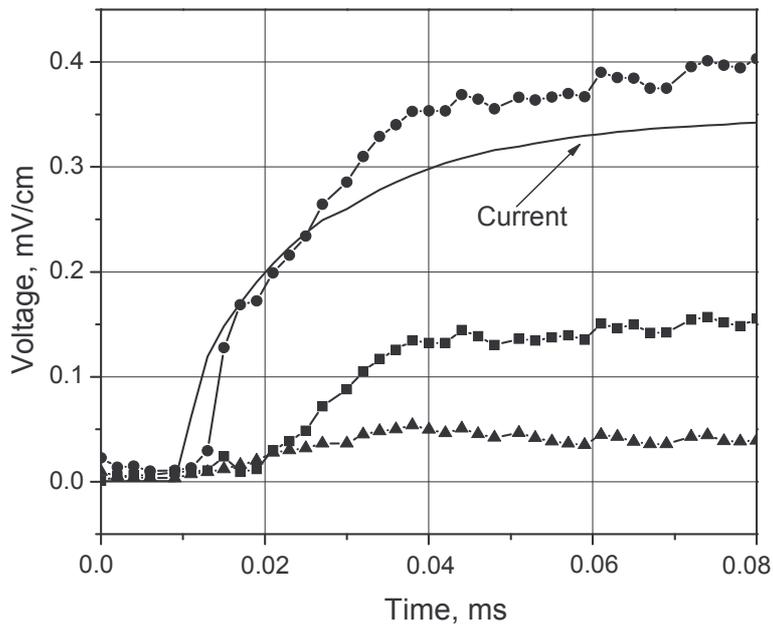

Fig. 5. Voltage traces for different amplitudes of the stepwise current after data processing procedure. The current trace is shown in arbitrary units.

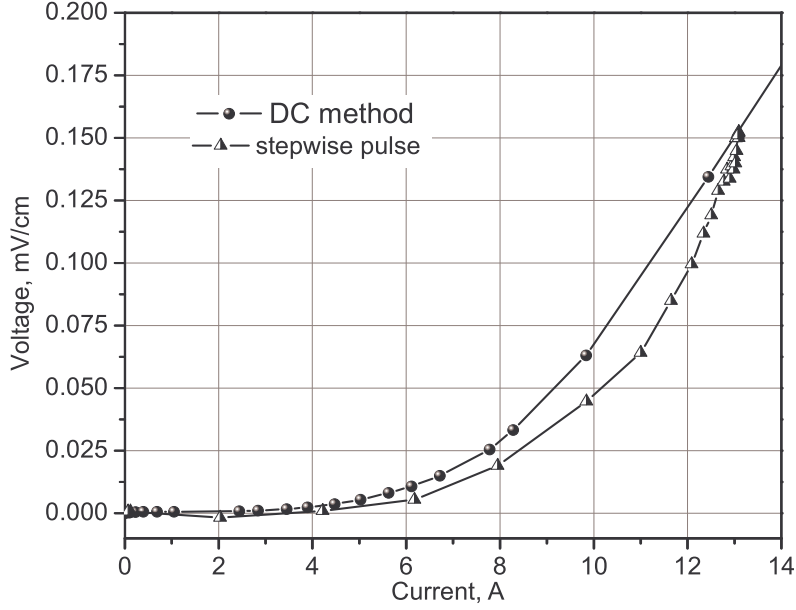

Fig. 6. Comparison of the dynamic and static VCCs. Dynamic VCC was obtained from the response of the central pair of taps to a stepwise current.

## 3. Theoretical analysis and computer simulation

### 3.1. Mathematical model

To understand the origin of observed peculiarities of the dynamic VCCs, we analyzed the non-linear magnetic diffusion into a thin tape theoretically. Let us consider a conducting strip in a zero external magnetic field. The strip is infinite along the *x* axis, has the width 2*L* along the *y* axis and the thickness 2*a*, $a \ll L$. Let a transport current $I(t)$ be applied. The electric field *E* and sheet current density $J(y, t)$ have only the *x*-components. At $z = 0$, the *z*-component of magnetic field is given by the Ampere law as

$$H_z(y,t) = \frac{1}{2\pi} \int_{-L}^{L} \frac{J(u,t)du}{y-u} \quad (1)$$

The total current is given by

$$I(t) = \int_{-L}^{L} J(y,t)dy. \quad (2)$$

In the framework of Bean's model, the current density in a superconductor is zero or equal to the critical value $|j| = j_c$ and the current distribution is formed without any delay correspondingly to instantaneous values of the external magnetic field and transport current.

Substituting Eq. (1) into the Faraday law equation

$$\frac{\partial E}{\partial y} = \mu_0 \frac{\partial H}{\partial t}, \quad (3)$$

where $\mu_0$ is the vacuum magnetic permeability, we obtain

$$\frac{\partial E}{\partial y} = \frac{\mu_0}{2\pi} \int_{-L}^{L} \frac{\partial J(u,t)}{\partial t} \frac{du}{y-u} \quad (4)$$

We use Eq. (4) to evaluate the diffusion time for various forms of the VCC of a homogeneous strip. For the linear *E-J* characteristic (Ohm law), $E = \rho_m J/2a$, where $\rho_m$ is a constant resistivity, Eq. (4) can be presented in the dimensionless form:

$$\frac{\partial e}{\partial \xi} = \frac{\mu_0 L a}{\pi \tau \rho_m} \int_{-1}^{1} \frac{\partial e(u,\vartheta)}{\partial \vartheta} \frac{du}{\xi - u} \quad (5)$$

where $e = E/E_0$, $\xi = y/L$, $\vartheta = t/\tau$, and the normalization parameter $E_0$ can be arbitrarily chosen. The characteristic time $\tau$ is

$$\tau = \frac{\mu_0 L a}{\pi \rho_m}. \quad (6)$$

Expression (6) can be also used to estimate a characteristic time of the magnetic diffusion into a superconductor shunted by a well conducting metal (coated superconductors, tapes, superconducting thin films covered by metal layer). In the case of a composite superconductor, in Eq. (6), the thickness and the resistivity of a superconductor should be replaced by the thickness and the resistivity of a normal conducting layer. For a superconductor described by the classical critical state model, the diffusion time is determined by a normal conducting layer only.

Suppose now the strip is characterized by a power law (flux creep regime) $E = E_c(J/J_c)^n$, where $J_c$ is the critical sheet current density; $J = 2aj$ and $j$ is independent of $z$. To rewrite Eq. (3) in a dimensionless form, we define the scaling sheet current density $J_0 = I_0/2L$ and the corresponding electric field $E_0 = E_c(J_0/J_c)^n$. Then Eq. (5) can be rewritten as

$$\frac{\partial e}{\partial \xi} = \frac{\mu_0 J_0 L}{2\pi n \tau E_0} \int_{-1}^{1} e^{1/n-1} \frac{\partial e(u,\vartheta)}{\partial \vartheta} \frac{du}{\xi - u}, \quad (7)$$

As an estimate for the characteristic diffusion time, one can choose

$$\tau = \frac{\mu_0 J_0 L}{2\pi n E_0} = \frac{\mu_0 I_0}{4\pi n E_0}. \quad (8)$$

Since the $E$-$J$ characteristic is non-linear, the diffusion time depends on the current pulse amplitude $I_0$. The parameter $E_0$ in (8) can be taken from the DC VCC as the electric field corresponding to the current amplitude. For a superconductor described by the power law VCC and shunted by a normal conducting metal, the characteristic time can be estimated as the largest value from the times determined from (6) and (8).

Let us note some differences of the expressions (6) and (8) from the obtained for slabs in a magnetic field parallel to their surface [19, 20]. The characteristic magnetic diffusion time decreases as the slab thickness is reduced, and dynamic hysteresis loops disappear in thin slabs. In the case of the linear $E$-$J$ characteristic, the diffusion time into a tape is determined by the product of its thickness and width (Eq. (6)) while for a slab in parallel field this time is proportional to the thickness squared. For non-linear characteristics, the diffusion time into a tape is determined by the VCC and not by the tape thickness (see Eq. (8)); the time is finite even for infinitely thin films and the loops do not disappear.

*3.2. Numerical algorithm*

The calculations were carried out for a superconducting composite similar to the tested tape (Fig. 1) with a superconductor inside an Ag matrix and occupying the central part of the tape, $|y| \leq l$ with different values of the ratio $l/L$. To develop a numerical algorithm, we integrate Eq. (4) with respect to $y$, substitute $E = \rho(y,J)J/2a$, where $\rho(y,J)$ is a non-linear resistivity corresponding to the $E$-$J$ characteristic of a superconductor, and obtain:

$$\rho(y,J)J = \frac{\mu_0 a}{\pi} \int_{-L}^{L} \frac{\partial J(u,t)}{\partial t} \ln|y - u| du + C(t) \quad (9)$$

where $C(t)$ is unknown function of time.

The system of Eqs. (2) and (9), determining both the current distribution in a tape and the function $C(t)$, was solved numerically.

Let $\{y_i\}$, $i = 0,...,N$, be an equidistant mesh on $[-L, L]$ with $\Delta = y_i - y_{i-1}$. We use piecewise-constant finite elements and approximate $J(y,t)$ by $\sum_{j=1}^{N} J_j(t)\varphi_j(y)$, where

$$\varphi_j(y) = \begin{cases} 1 & y_{j-1} \leq y \leq y_j, \\ 0 & otherwise \end{cases}$$

Multiplying equation (9) by $\varphi_i$, $i = 0,...,N$, and integrating, we obtain

$$M\dot{\bar{J}} = \Delta\left(\bar{b}(\bar{J},t) - C\bar{1}\right)$$

where $b_i(\bar{J},t) = \rho_i(J_i)J_i$, $\rho_i(J_i) = \rho(y_{i-1/2}, J_i)$, $y_{i-1/2} = y_i - \Delta/2$, $\bar{1} = (1,...,1)^T$, a point above a symbol notes the time derivative. The integrals $M_{ij} = \frac{1}{2\pi}\int_{-L}^{L}\int_{-L}^{L}\varphi_j(x)\varphi_i(y)\ln|x-y|dxdy$ can be found analytically: $M_{ij} = -\frac{\Delta^2}{4\pi}(m_{|i-j|} + 2\ln\Delta)$ with

$$m_k = \begin{cases} -3 & k=0, \\ -3 + 4\ln 2 & k=1, \\ -3 + (k+1)^2\ln(k+1) - 2k^2\ln k + (k-1)^2\ln(k-1) & k \geq 2. \end{cases}$$

For each time moment, we substitute $\dot{\bar{J}} = \Delta M^{-1}(\bar{b}(\bar{J},t) - C\bar{1})$ into the equation $\dot{I}(t) = \Delta\sum_{i=1}^{N}\dot{J}_i$, obtained from (2), and find

$$C = \left(\sum_{j=1}^{N}(M^{-1}\bar{b}(\bar{J},t))_j - \frac{\dot{I}}{\Delta^2}\right) \bigg/ \sum_{i,j=1}^{N}M^{-1}_{ij}.$$

Now, using a standard routine for ordinary differential equations, we can integrate over time the system

$$\dot{\bar{J}} = \Delta M^{-1}(\bar{b}(\bar{J},t) - C\bar{1})$$

with zero initial conditions. Note that the inverse matrix $M^{-1}$ should be computed only once.

*3.3. Simulation: comparison with experiment*

The simulation was performed for a composite monofilamentary tape shown in Fig. 1 with the *E-J* characteristic taken in the form:

$$E = \begin{cases} \rho_m J/2a & \text{for } -L \leq y \leq -l \\ E_0(J/J_c)^n & \text{for } -l < y < l \\ \rho_m J/2a & \text{for } l \leq y \leq L \end{cases} \quad (10)$$

where $\rho_m$ is now related to the resistivity of the matrix.

Expressions (6) and (8) cannot be directly applied to a superconducting composite. For example, for Bean's model ($n\to\infty$) Eq. (8) correctly gives zero diffusion time for a superconductor, in this case the diffusion time for a composite tape is mainly determined by diffusion in a normal matrix. In calculations we used the dimensionless units with the time scale $\tau$ chosen for a normal metal tape according to Eq. (6). In these units, the resistivity of the normal conducting parts near the tape edges, $l < |y| \leq L$, is equal to 1. For the central part, $|y| \leq l$, containing the superconductor, we used the power law characteristic (flux creep model) with the resistivity corresponding to the experimental data, $\rho(J) = 0.015 |J/J_c|^3$ ($n = 4$). For the tested tapes, the critical current $I_c = 2lJ_c$ is about 4.6 A and Eq. (6) gives the characteristic time of about $\tau = 2.5\cdot10^{-5}$ s. The geometric parameters in dimensionless units are $L = 1$, $l = 0.8 \div 0.95$.

In order to simulate the experimental conditions, the sinusoidal pulse was taken in the form $I = I_0 \sin(\pi t/36)$ where $I_0$ is the pulse amplitude in the units of $I_c$ and the time is taken in units of $\tau$. The stepwise pulse was described by the expression $I = I_0\{1-exp(-t/0.8)\}$. To compare the simulation and experimental results, we applied to the calculated curves the same subtraction procedure as was used for processing of the experimental data (see Section 2.3).

The simulation results for both sinusoidal and stepwise pulses are presented in Figs. 7-9. They are in qualitative agreement with our experimental data. The calculated curves have the same features that were observed in the experiments. For the sinusoidal pulse, the simulation gave the loops similar to the experimental ones (Fig. 7): broad clockwise loops at the edge (Fig. 7a) and narrow anticlockwise loops at the center (Fig. 7b). Some difference between the experimental and calculated results can be explained by finite size of the potential taps which cannot be made as points, inhomogeneities of superconducting properties of the samples, a finite contact resistance between the matrix and the superconductor.

Note that, as opposed to the experiment, the calculated voltage curves do not contain any components related to measuring circuit. Therefore, the calculated curves show us a "pure" voltage drop across the tape (Figs. 7a,b, 8a, 9). This voltage drop contains a component induced inside the tape, non-linear in current amplitude. This component determines the difference of the dynamic VCCs measured at the center and at the edge. The subtraction procedure applied to the calculated curves removes the main part of this component (Figs. 7c,d, 8b, 9). However, a part of this component remains. Therefore, the dynamic VCCs for the edge points go above the DC curve in the lower part. The subtraction removes also a part of the active component. This leads to lower voltage in the higher part of the dynamic curve. The influence of the subtraction procedure manifests itself especially clearly for the stepwise pulse (Figs. 8 and 9).

Note that after the subtraction we have got self-intersecting loops as after the data processing of the experimental data (compare Figs. 3a and 7c).

Increasing step by step the pulse amplitude, we obtain a set of similar loops embedded one inside the other (such loops were also obtained experimentally). The main difference between all the obtained VCCs is observed in the range of low voltages where a criterion for the critical current value is usually chosen (1-10 µV/cm). Therefore, the determination of the critical current from the dynamic VCC becomes to be problematic: the obtained value depends on location of the measurement taps and on the branch of the dynamic curve (ascending or descending). For example, if to determine the critical current from the ascending branch at the center (Fig. 7d), we obtain the critical current increasing with the pulse amplitude or, that is the same, with the rise rate of the current. The dependence of the critical current obtained from the descending branch becomes opposite: it decreases with the increase of the pulse amplitude.

*3.4. Simulation: comparison between flux creep and Bean's models*

Simulating the magnetic diffusion into a thin tape in the framework of the flux creep approach (power-law characteristic (10) for the superconductor), we have obtained a good agreement with the experiment. To understand clearly the influence of the flux creep in a superconductor on the magnetic field diffusion in a composite, let us compare our results with those obtained using Bean's critical state model [31]. In the framework of Bean's approach, the diffusion coefficient approaches infinity and the magnetic field distribution inside a superconductor is adjusted instantly to an applied field or a current. The calculations were performed for Bean's model modified to account for the presence of a normal matrix: $\rho(J)J = 0$ at a sheet current less than the critical value $J_c$, and $\rho(J)J = \rho_m \text{sign}(J)(|J|-J_c)$ for $|J| > J_c$ at $|y| \leq l$. Fig. 10 shows the calculated VCCs for the tape of Fig.1 subjected to the sinusoidal pulse. Bean's model gives also the loops with opposite path-tracing for central and edge-assigned potential taps (compare with Fig.7). For both models we obtain broad clockwise loops at the edge and narrow anticlockwise loops at the tape centre. In the range of low currents, voltage at the edge is determined by the magnetic diffusion into a normal conducting matrix. Therefore, the loops at the edge practically coincide for both models. Figs. 11 and 12 show the penetration of the current and electric field into a tape for flux creep and Bean's models. The process includes two stages. First, the electric field front propagates inside the tape with the velocity determined by the rise rate of the applied current and the magnetic diffusion into the normal matrix. For the considered case this stage takes about $7\tau$. For Bean's model (Fig. 11), this time corresponds to appearance of the electric field at the center (for the considered case, at the current of about 1.1 $I_c$), and after that the practically uniform distribution is established.

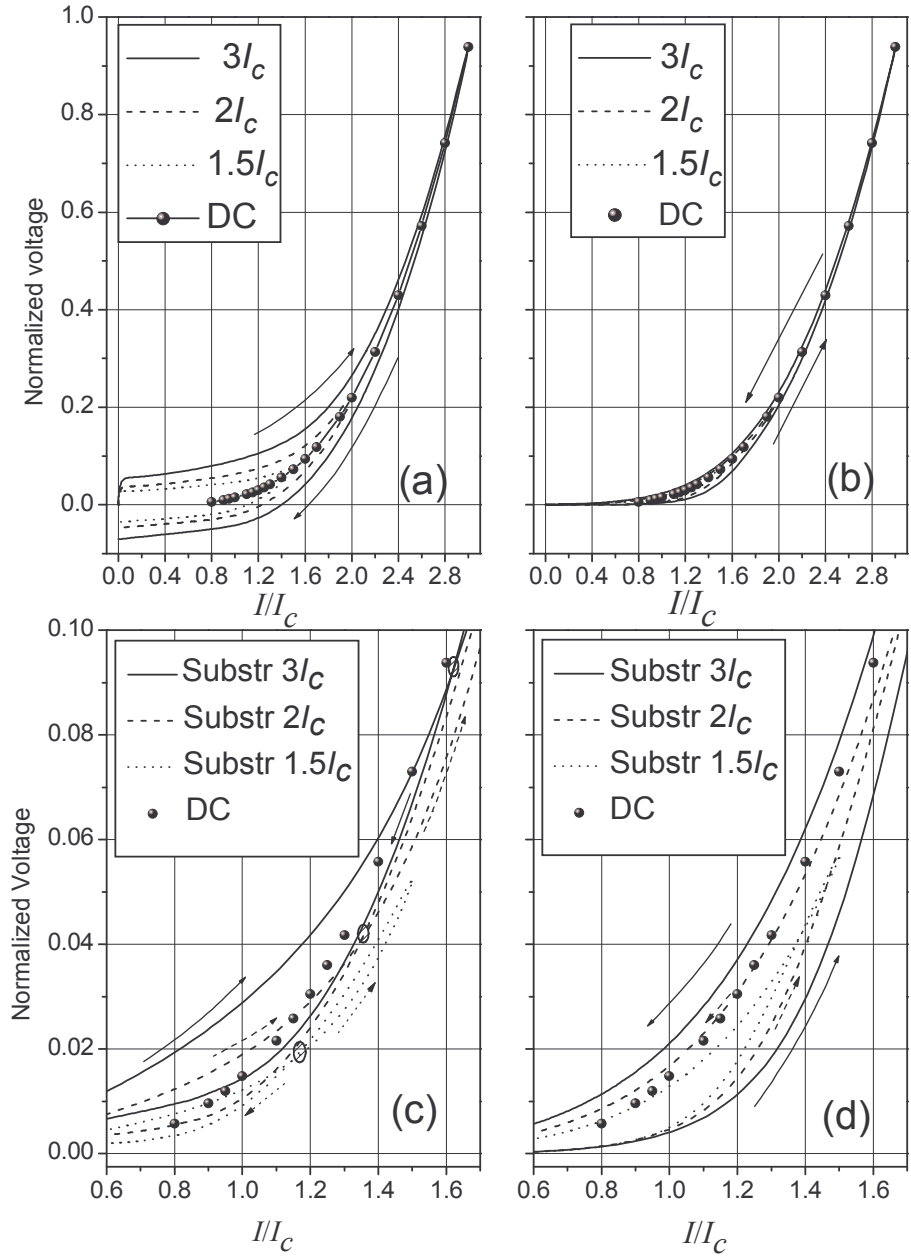

Fig. 7. Simulation of the dynamic VCCs of a tape with the power-law characteristic for a half-wave of the sinusoidal pulse $I = I_0 \sin(\pi t/36)$ with different amplitudes $I_0$. Shown: calculated voltages at the edge (a) and at the center (b) before the subtraction procedure; (c) and (d) are the same after the subtraction. The arrows indicate the path tracing.

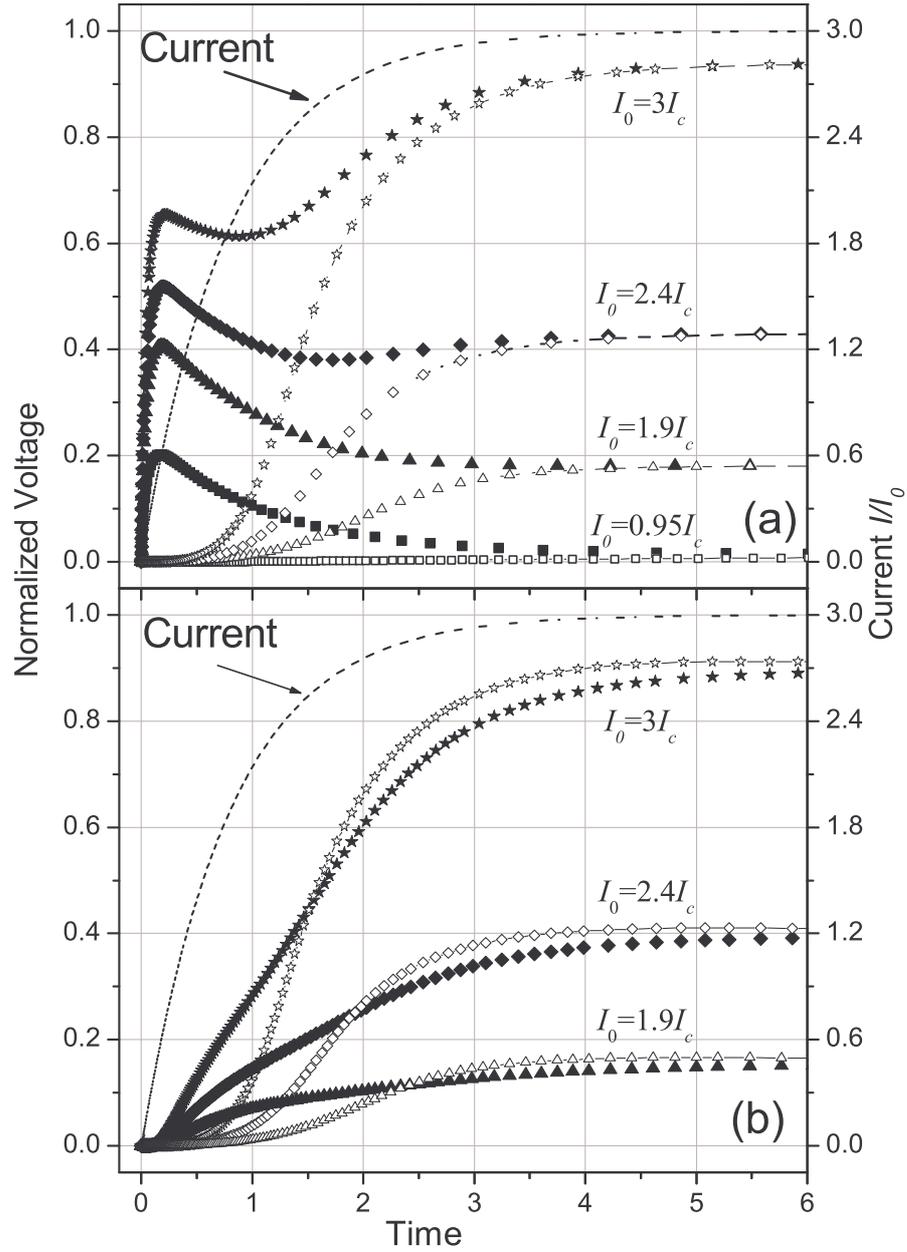

Fig. 8. Simulation of the dynamic behavior of the tape subjected to the current $I = I_0\{1-exp(-t/0.8)\}$: (a) voltage drops at the edge and at the center before the subtraction procedure; (b) the same after the subtraction procedure.

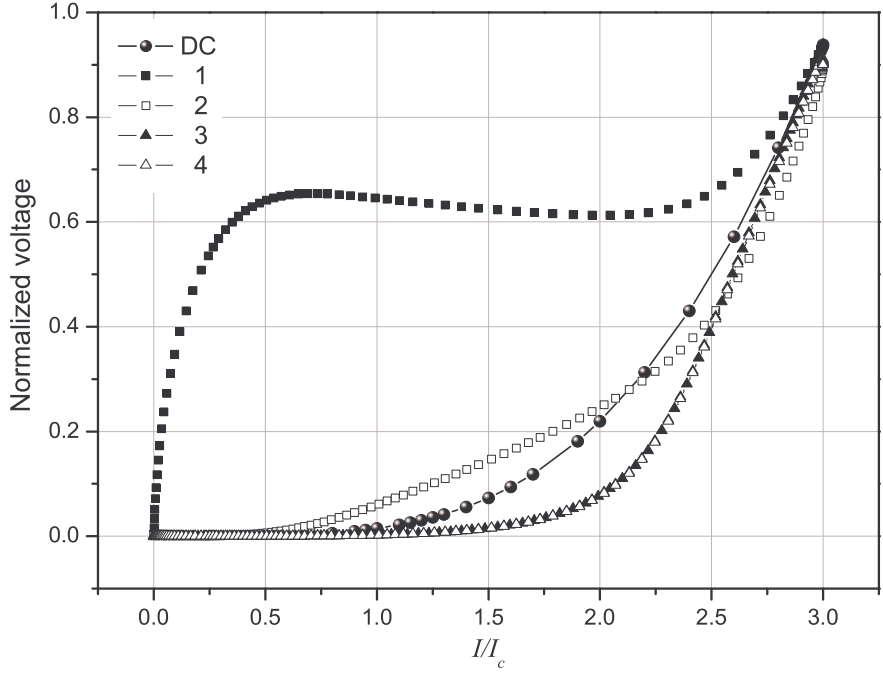

Fig. 9. Calculated dynamic VCCs versus calculated DC VCC: shown the voltage at the edge (curves 1 and 3) and at the center (curves 2 and 4) before and after the subtraction of induced components, respectively.

For the flux creep model (Fig. 12), we observe clearly when the electric field the non-linear magnetic diffusion prevails. Note, that in contrast to Bean's model, the power law VCC gives a nonzero electric field at the center at any current value. The difference between the diffusion characters in both cases is illustrated by the simulation presented in Fig. 13: shown the response of the tape to the application of a stepwise pulse with a very short front. As opposed to Bean's behavior, the diffusion time for the flux creep depends on the pulse amplitude in full accordance with (8). Note that for modern HTS composites with high critical currents, the time of the magnetic diffusion can exceed sufficiently the time observed in our experiments. For example, for $I_c = 100A$ and the pulse amplitude $I_0 = 1.5\ I_c$ this time becomes equal to 7.5 ms.

## 4. Conclusion
The magnetic flux penetration and the dynamic VCCs of Bi-2223/Ag tapes have been studied both experimentally and theoretically.
1. The dynamic VCCs deviate substantially from the DC VCC. The value and sign of the deviations depend on the location of potential taps at the surface of the sample: there is the large difference between the signals on the edge- and center-assigned taps. The subtraction of the inductive component, ever appearing in the measurements, can lead to some distortions due to the non-linear E-J characteristic of the sample – the fact which should be taken into account at the analysis of the results.
2. The clockwise and anticlockwise hysteresis loops in dynamic VCCs can be observed for the same sample depending on location of the potential taps. This is a specific feature of tapes, films and other samples with a planar geometry.
3. It was shown that the observed peculiarities of the dynamic characteristics of monofilamentary superconducting tapes can be explained in the framework of the magnetic diffusion and flux creep mechanism.

4. It is essential that the sufficient deviations from the DC VCC are observed for relatively long pulses with the rise time 20 times exceeding the characteristic time of the magnetic diffusion determined in the framework of Bean's model. Deviations from the DC values are especially noticeable in the range of low voltages, the more important zone for the critical current determination. The measured critical current and its change with the pulse amplitude depend on the location of the potential taps and on the VCC branch chosen for the critical current determination. More accurate measurement of VCCs and critical currents can be obtained with center-assigned taps.

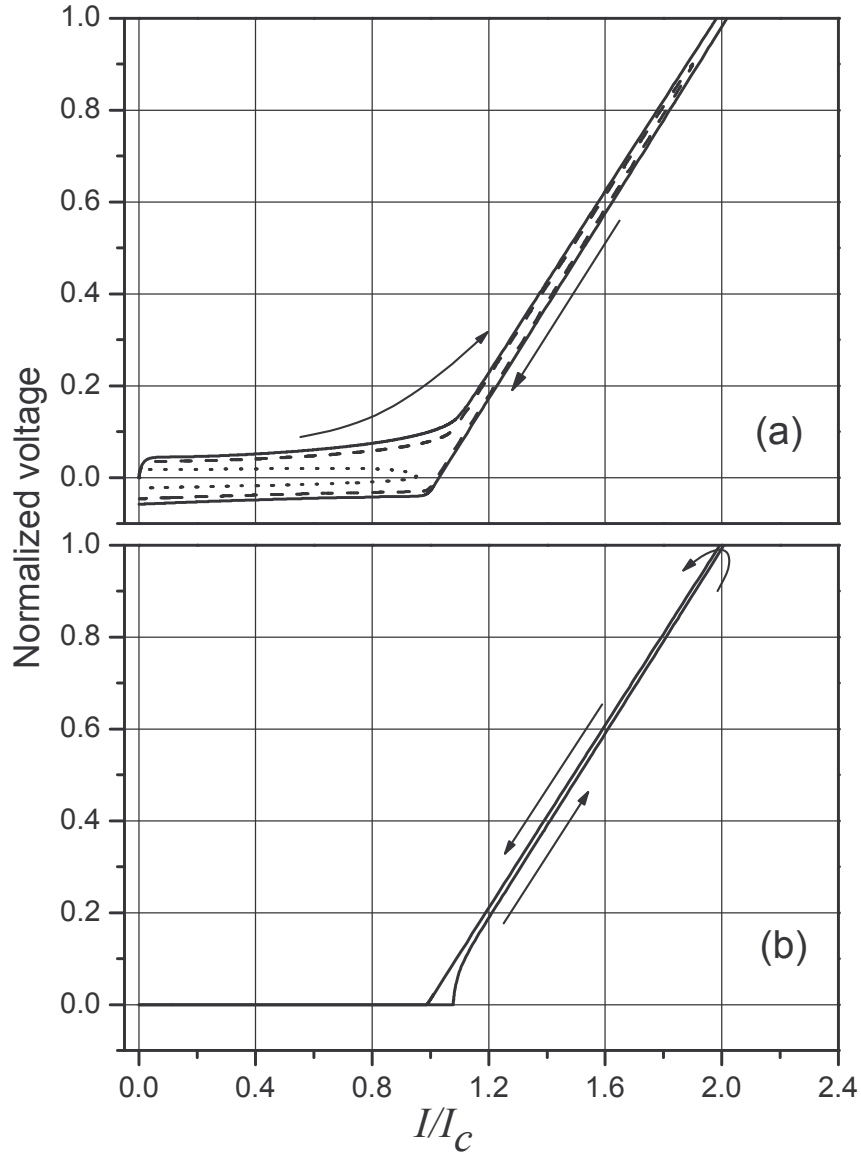

Fig. 10. Simulation of the dynamic VCCs of a tape with Bean's characteristic for a half-wave of the sinusoidal pulse $I = I_0 \sin(\pi t/36)$ with different amplitudes $I_0$. Calculated voltages at the edge (a) and at the center (b) are shown. The arrows indicate the path tracing.

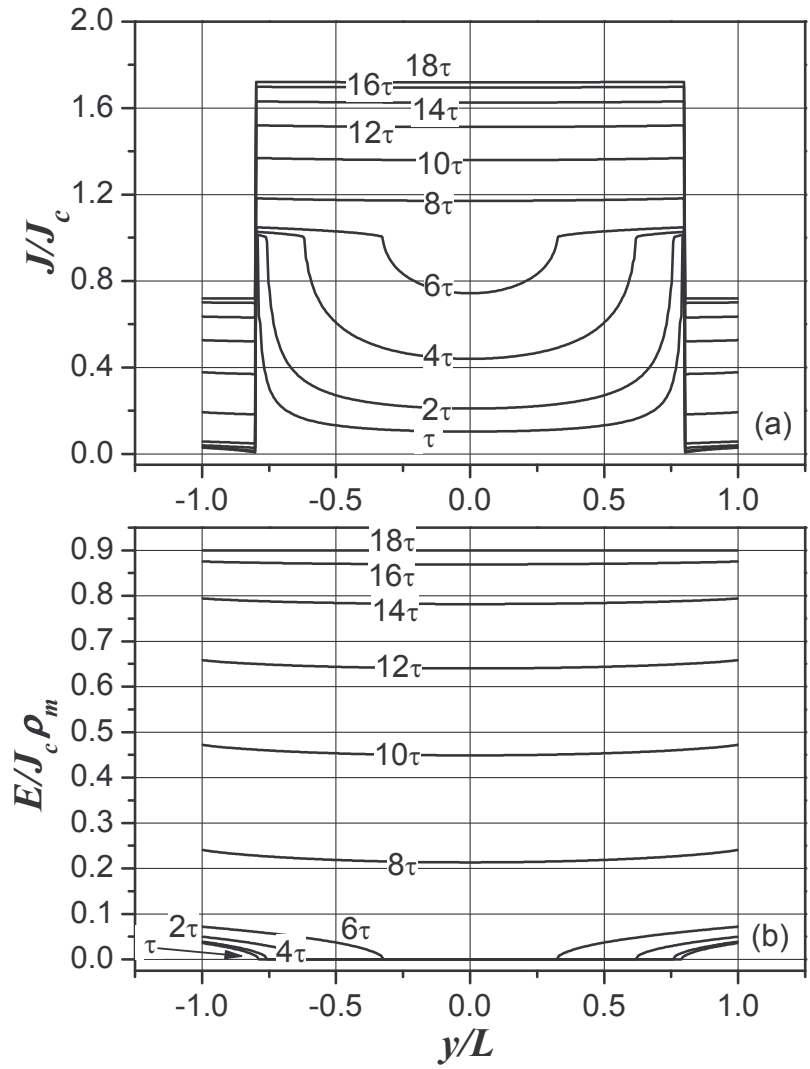

Fig. 11. Calculated current (a) and electric field (b) distributions in a tape for Bean's model. Current pulse $I = I_0 \sin(\pi t/36)$ with the amplitude $I_0 = 1.9\, I_c$.

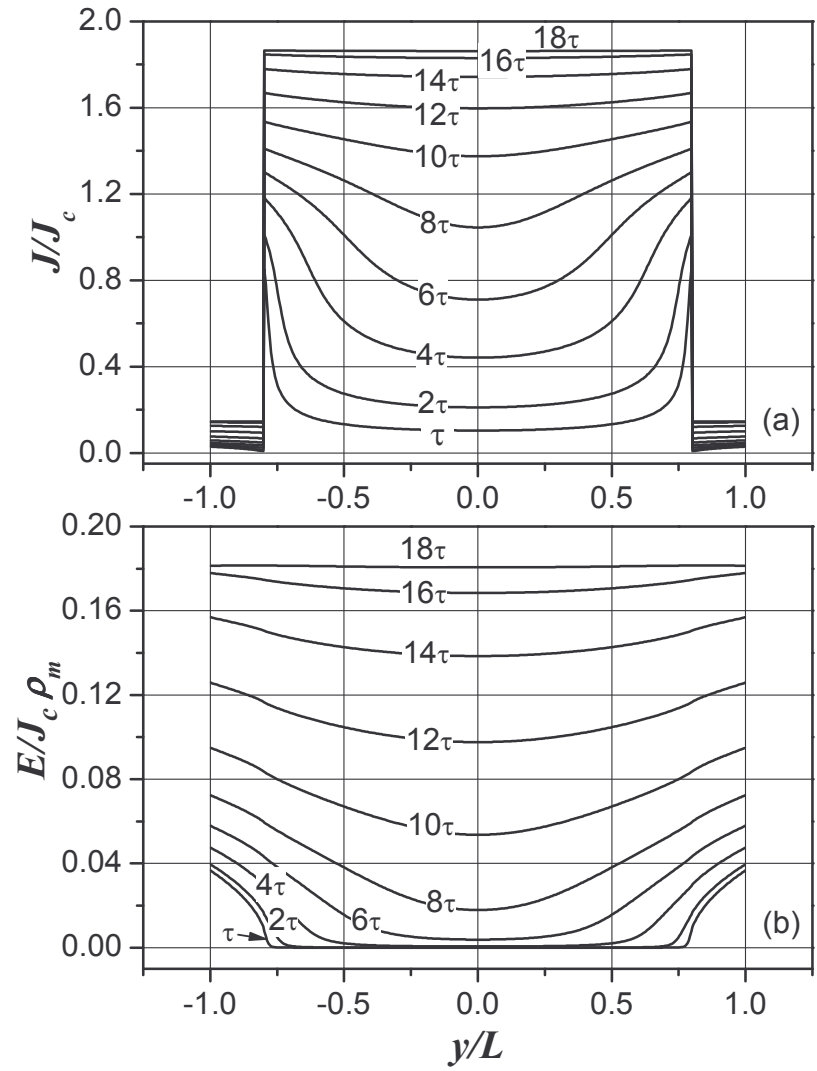

Fig. 12. Calculated current (a) and electric field (b) distributions in a tape for the flux creep model. Current pulse $I = I_0 \sin(\pi t/36)$ with the amplitude $I_0 = 1.9\, I_c$.

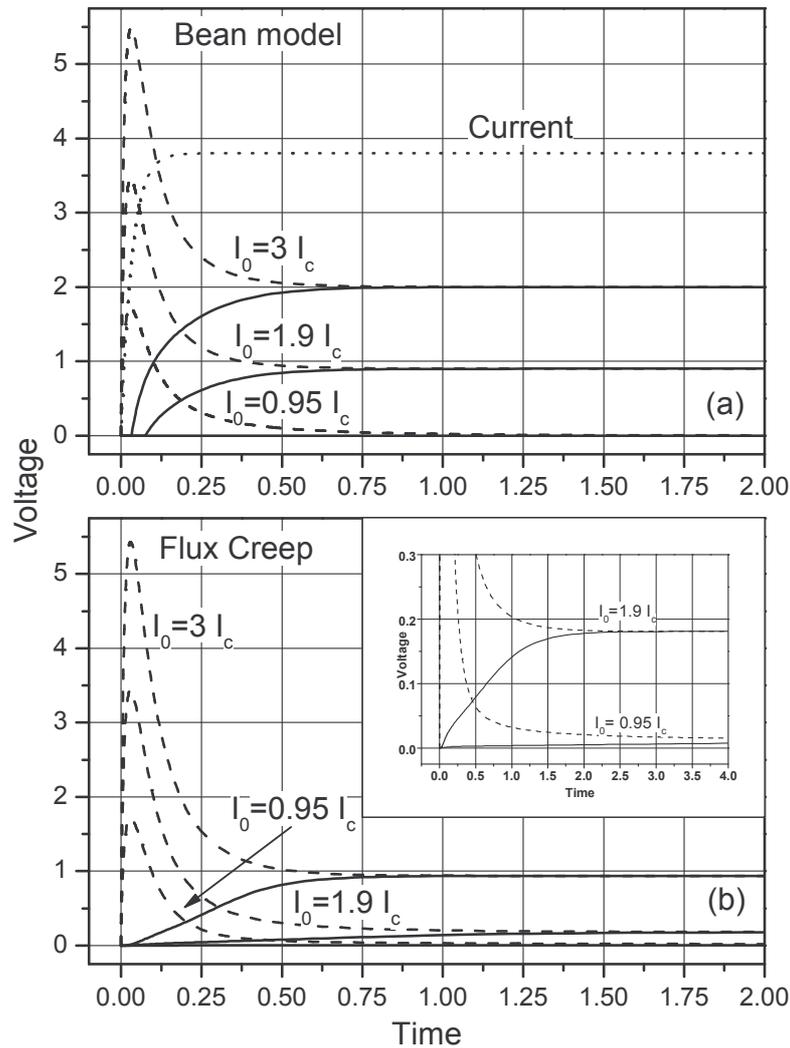

Fig. 13. Simulation of the dynamic behavior of the tape subjected to a stepwise current pulse with short rise time $I = I_0\{1-exp(-30t)\}$ for Bean' model (a) and the flux creep model (b). Voltage drops at the edge (dashed curves) and at the center (solid curves) are shown for different $I_0$.

**Acknowledgments**
The authors wish to thank Dr. Beilin, Hebrew University, Israel, for fabrication of the experimental samples.